\documentclass[sigconf]{acmart}

\AtBeginDocument{%
  \providecommand\BibTeX{{%
    \normalfont B\kern-0.5em{\scshape i\kern-0.25em b}\kern-0.8em\TeX}}}

\acmConference[MSR 2022]{MSR '22: Proceedings of the 19th International Conference on Mining Software Repositories}{May 23–24, 2022}{Pittsburgh, PA, USA}
\usepackage{listings}
\usepackage{acronym}
\usepackage{hyperref}

\acrodef{LAGOON}{Leveraging AI to Guard Online Open Source Networks}
\acrodef{PEP}{Python Enhancement Proposal}
\acrodef{NLP}{natural language processing}
\acrodef{OSS}{open source software}
\acrodef{UI}{user interface}
\acrodef{CLI}{command line interface}
\acrodef{ML}{machine learning}
\acrodef{GCN}{graph convolutional network}
\acrodef{OCEAN}{Open-source Complex Ecosystem And Networks}

\newcommand{\code}[1]{\texttt{\textcolor{blue}{#1}}}
\newcommand{\blueurl}[1]{\textcolor{blue}{\url{#1}}}

\begin{document}
\sloppy

\title{{LAGOON}: An Analysis Tool for Open Source Communities}

\author{Sourya Dey}
\authornote{Both authors contributed equally to this work.}
\email{sourya@galois.com}
\orcid{0000-0003-3084-1428}
\author{Walt Woods}
\authornotemark[1]
\email{waltw@galois.com}
\orcid{0000-0001-8489-9436}
\affiliation{%
  \institution{Galois, Inc.}
  \streetaddress{421 SW 6th Ave., Ste.\ 300\\}
  \city{Portland}
  \state{Oregon}
  \country{USA}
  \postcode{97204}
}

\renewcommand{\shortauthors}{Dey and Woods}

\begin{abstract}
This paper presents LAGOON -- an open source platform for understanding the complex ecosystems of Open Source Software (OSS) communities. The platform currently utilizes spatiotemporal graphs to store and investigate the artifacts produced by these communities, and help analysts identify bad actors who might compromise an OSS project's security. LAGOON provides ingest of artifacts from several common sources, including source code repositories, issue trackers, mailing lists and scraping content from project websites. Ingestion utilizes a modular architecture, which supports incremental updates from data sources and provides a generic identity fusion process that can recognize the same community members across disparate accounts. A user interface is provided for visualization and exploration of an OSS project's complete sociotechnical graph. Scripts are provided for applying machine learning to identify patterns within the data. While current focus is on the identification of bad actors in the Python community, the platform's reusability makes it easily extensible with new data and analyses, paving the way for LAGOON to become a comprehensive means of assessing various OSS-based projects and their communities.
\end{abstract}



\begin{teaserfigure}
  \includegraphics[width=\textwidth]{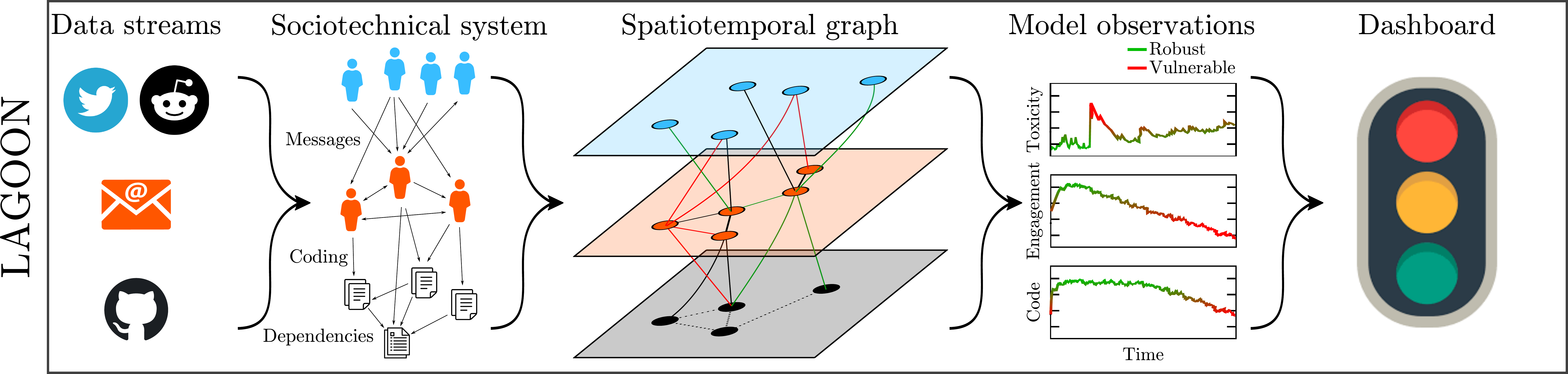}
  \caption{The \acs{LAGOON} platform.}
  \Description{The LAGOON platform. Data streams form a sociotechnical system, which gets ingested into our tool to form a spatiotemporal graph. Subsequently, observations can be drawn.}
  \label{fig:teaser}
\end{teaserfigure}

\maketitle

\section{Introduction}\label{sec:intro}
\Ac{OSS} is responsible for many of the productivity advancements in modern software engineering, essentially by expanding the author list of any software project considerably. However, \ac{OSS} can also directly cause failures in the projects that rely on them, with examples such as the unpublishing of \texttt{Leftpad} \cite{leftpad2018}, the corruption of \texttt{Faker.js} \cite{squires2022} and the pushing of malicious code to \texttt{UA-Parser.js} \cite{uaParser2021}. Understanding what motivates the maintainers of high-quality \ac{OSS} projects -- and being able to identify when such a project might become derelict or compromised -- is paramount to avoiding this class of supply-chain disruptions, which at best affects service uptime or requires considerable developer hours to remedy, and at worst can introduce subtle security bugs that compromise users.

On one hand, \ac{OSS} communities produce many artifacts, such as source code or mailing list messages, that can help understand the state of each project; on the other hand, such analysis is difficult due to its scope. Successful \ac{OSS} communities tend to be large, distributed and highly interconnected. Two issues further complicate matters: (1) the list of authors actively contributing to an \ac{OSS} project changes over time, which needs to be taken into account to contextualize a community's dynamics \cite{Klug2014,Oettershagen2020,Gionis2016}; and (2) many community members make significant contributions outside of the code base, such as ideations and discussions, requiring an analysis platform capable of ingesting and merging multiple, disparate data sources. Available tools to help with this analysis are limited.

We present the \acf{LAGOON} platform -- an open source, reusable tool to ingest, analyze and visualize temporal, sociotechnical graph data, in a way that allows for the assessment and analysis of \ac{OSS} projects and communities. Specifically, \ac{LAGOON}'s analyses are focused on detecting cyber-social operations targeting \ac{OSS} developer communities, though its modular design allows for additional use cases. \Ac{LAGOON} can currently be pointed at arbitrary \ac{OSS} code repositories or other sources of information, and users can create and add their own ingest modules or analyses. \Ac{LAGOON} is available at \blueurl{https://github.com/GaloisInc/SocialCyberLAGOON}, and can be cited as \cite{lagoon}. The URL includes a link to a publicly available backup of our database for the CPython project and its associated community artifacts, which form the primary case study for our analyses.

Notably, \ac{LAGOON} is a partner project to the \ac{OCEAN} data collection effort \cite{uvm_ocean} -- a partnership between Google Open Source and the University of Vermont. \Ac{OCEAN} provides standardized data from mailing lists of several \ac{OSS} communities, including $954,287$ messages from the Python community, which formed a key part of the data we ingested and analyzed in \ac{LAGOON}.


\subsection{Overview of \acs{LAGOON}}
Fig. \ref{fig:teaser} gives an overview of the \ac{LAGOON} platform. For a complete view of a sociotechnical system such as an \ac{OSS} community, \ac{LAGOON} supports the ingest and combining of data streams from multiple different sources representing the system into a spatiotemporal graph database, which is used for analyses.

Disparate sources are handled via an \textit{entity} and \textit{observation} model. That is, each unique entity is created as a node in the graph; entities are intransient. At different points in time, two entities might interact or a single entity might be measured in some way: these are recorded into the graph as observations, and always have timestamps (though NULL might be used to indicate an intransient observation). For example, Person A (an entity) might author on July 2nd, 2020 (an observation) Message B (another entity). Both entities and observations can have arbitrary data attached for convenience; e.g., Person entities have a username and e-mail address.

Entities and observations are ingested from each data source as part of a \textit{batch}, independently from all other data sources. That is, ingesting a git repository creates a set of user entities which are completely separate from the user entities produced during a mailing list ingest. This allows for data sources to be individually re-ingested or updated to include new features without requiring the entire database to be re-ingested. 

Dealing with duplicate or aliased entities from disparate sources of data has been discussed in literature \cite{Enamorado2019,Fry2020,Wiese2016}. As the final step of database preparation, \ac{LAGOON} ingest batches are merged under a \textit{fusion} process that de-duplicates entities across different batches by creating a single \textit{FusedEntity} to represent them.

The end result is a database ready for analysis, providing a unified view of the disparate data sources created by \ac{OSS} communities.

\ac{LAGOON} contains a Python library for querying the database. This library has been used to derive all of our results, as well as some results for \ac{OCEAN}. \Ac{LAGOON} is also packaged with a \ac{UI} for inspecting and analyzing the graph, as well as a \ac{CLI} for ad hoc queries.


\subsection{Related work}
The popularity of the graph format have led to a profusion of graph datasets for \ac{ML} purposes \cite{Hu2021,Morris2020,Rozemberczki2020,Husain2020,icij_offshoreleaks}, some of which deal with code analysis and \ac{OSS} communities \cite{Husain2020,Rozemberczki2020}. We make the key distinction that these are datasets, not tools. In the tool space, there are several commercial / enterprise software for dealing with graphs \cite{anzo,tigergraph,hume,graphistry}, as well as open source efforts with predominantly visualization functionality \cite{franz2016cytoscape,neovis,neodash}, or analysis functionality \cite{graph-tool}. \ac{LAGOON}, as already described, is an open source tool which can ingest, visualize and analyze graph data. There have been efforts which have performed data analysis on \ac{OSS} data, such as \cite{Choudhary2018,Young2021,Ahmed2018}; these efforts do not include a comprehensive platform which also supports data ingest, and are in a similar vein to the results we present in Sec. \ref{sec:ana}.

\section{The \acs{LAGOON} Platform}\label{sec:lagoon}

\begin{figure*}[tbp]
  \centering
  \includegraphics[width=0.9\linewidth]{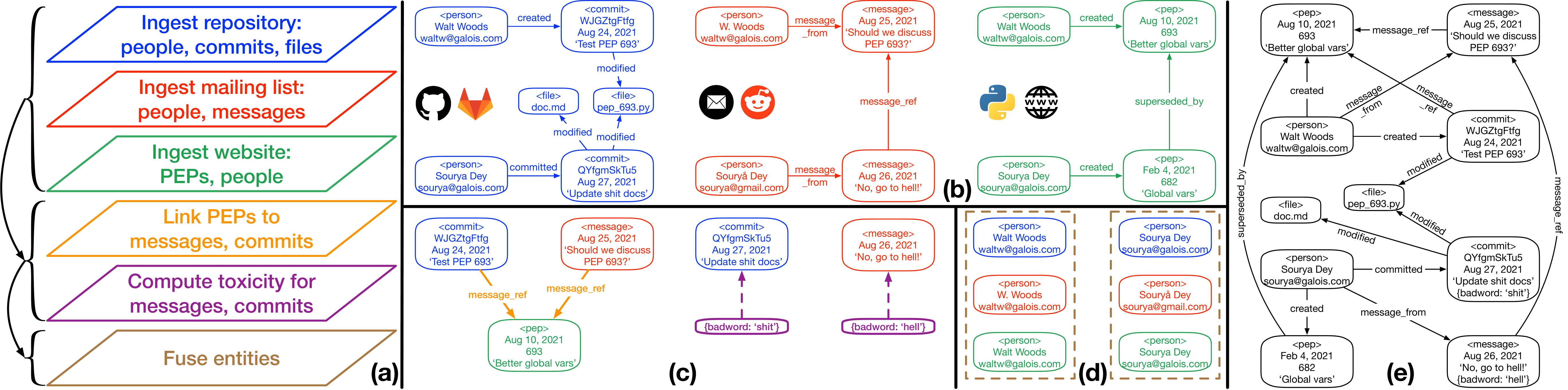}
  \caption{A complete example of \acs{LAGOON}'s different ingestion processes and fusion. Should be viewed in color. (a) Data sources. (b) Data is ingested from repositories (blue), mailing lists (red), and web scraping of \acsp{PEP} (green) in independent batches. (c) As separate batches, ingested messages are linked through observations to the \acsp{PEP} they reference (orange), and toxicity attributes are added to existing messages (purple). (d) Finally, person entities are fused (brown). (e) The final graph with all data co-existing. Note: Batches within each of the subfigures (b), (c), (d) can be reordered, but the ordering (b)$\rightarrow$(c)$\rightarrow$(d) is required.}
  \Description{LAGOON graph complete example}
  \label{fig:graph}
\end{figure*}

\ac{LAGOON} is a tool built using Python and contains a PostgreSQL database \cite{postgresql} accessed using SQLAlchemy \cite{sqlalchemy}. Fig. \ref{fig:graph} gives a detailed example of the platform ingesting data from an \ac{OSS} project. Data sources (a) are processed into batches (b), which can be cross-linked (c) and fused (d) to form a final sociotechnical graph (e). An example database is provided with \ac{LAGOON}, which contains artifacts relating to CPython, the reference implementation of Python.

\subsection{Ingest}
\ac{LAGOON} currently supports several ingestion modules. Users can also extend \ac{LAGOON} with their own custom ingestion modules.

\subsubsection{Ingesting from repositories}
As entities, this module pulls in (1) people (i.e. maintainers) with their name and email address; (2) commits with their hash, commit time and message; and (3) files within the repository. Observations are used to link maintainers to the commits they created, and commits to the files they changed.

Example: The blue batch in Fig. \ref{fig:graph}(b) ingests from a repository two maintainers -- Walt Woods and Sourya Dey -- each of whom has made one commit.

\subsubsection{Ingesting from mailing lists}
As entities, this module pulls in messages with their time, subject, body and source mailing list; and people (i.e. authors), as from repositories. Observations are used to link authors to messages, and messages referencing each other.

Example: The red batch in Fig. \ref{fig:graph}(b) ingests from a mailing list two authors -- W. Woods and Souryå Dey -- each of whom has written one message, with the latter's referencing the former's.

\subsubsection{Ingesting from web scrape -- \acsp{PEP}}
As entities, this module pulls in \acfp{PEP} \cite{peps} with their title, number and other information as obtained from the leading table of a \ac{PEP}'s webpage\footnote{See \blueurl{https://www.python.org/dev/peps/pep-0435/} for an example}; and people (i.e. \ac{PEP} authors), as described above. Observations link authors to \acp{PEP}, and associated \acp{PEP} together.

As a separate batch, this module links the ingested \acp{PEP} to previously ingested batches containing entities which mention that \ac{PEP} by number.

Example: The green batch in Fig. \ref{fig:graph}(b) ingests two \acp{PEP} -- 693 created by Walt Woods and 682 created by Sourya Dey, with \ac{PEP} 693 superseding \ac{PEP} 682. Then, the orange batch in Fig. \ref{fig:graph}(c) creates observations to link entities from previous batches to the \acp{PEP}. In this case, there was one commit message in the blue batch and one mailing list message in the red batch which contained `PEP 693'.

\subsubsection{Ingesting toxicity attributes}
This module analyzes messages from previous batches for toxic language, and attaches attributes to the respective entities. Our current database uses toxicity lists from \cite{Rezvan2018}. Toxicity analysis was separated from, e.g., mailing list ingest, to minimize the runtime of each separate step. 

Example: The purple batch in Fig. \ref{fig:graph}(c) identifies `shit' and `hell' as toxic words and attaches them to the respective entities -- one commit and one message. Note that this attaching is not done via creating observations (hence the arrows are not solid), instead the attributes dictionary of the respective entities are augmented.

\subsection{Entity fusion}
\ac{LAGOON} fuses person entities if any of these criteria are met:
\begin{itemize}
    \item Emails are identical and valid (i.e. not part of a generic mailing list address).
    \item Names are identical, and the names have a space in them and are longer than 5 characters.
    \item Names, on passing through PostgreSQL’s \texttt{fuzzystrmatch}’s metaphone function, have a similarity of $>0.95$ according to \texttt{pg\_trgm}’s similarity function. This allows for names with slight typos and/or missing accents to be merged.
\end{itemize}
Entity fusion finalizes the database by replacing entities and observations with \texttt{FusedEntity}-es and \texttt{FusedObservation}s. Our most recent CPython database contains about a million \texttt{FusedEntity}-es and 4 million \texttt{FusedObservation}s (as of 2022-01-24).

Example: Entities inside each brown box in Fig. \ref{fig:graph}(d) are fused. Note that `Sourya Dey' is fused with `Souryå Dey' since the two names are similar, and `Walt Woods' is fused with `W. Woods' since their emails are the same. The final data in the graph is shown in Fig. \ref{fig:graph}(e), which combines all the steps from Fig. \ref{fig:graph}(a).

\subsection{User Interface}
\ac{LAGOON} is packaged with a \ac{UI} for inspecting and analyzing the ingested open source community. The \ac{UI} was constructed with the Vue 3 framework \cite{vuejs}.

\begin{figure}[t]
    \centering
    \includegraphics[width=0.9\linewidth]{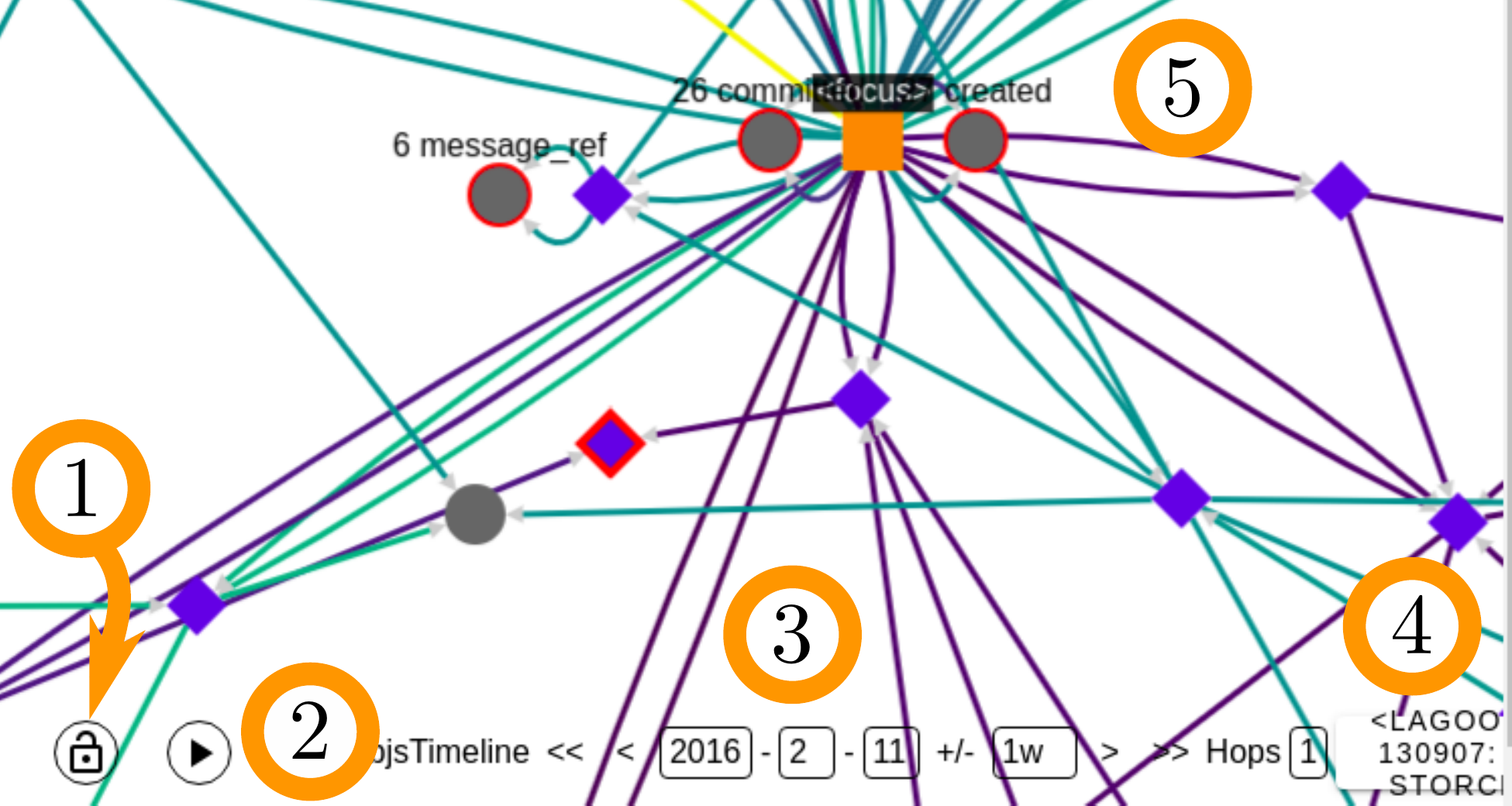}
    \caption{Small example screenshot from \acs{UI}, with key parts annotated by bold orange lines. (1) Focus locking, which can be used to quickly query displayed entities. (2) Pause/resume of continuous layout. (3) time range displayed from the database; observations at the start of this range are colored dark purple, while observations at the end of this range are bright yellow. (4) Name of currently focused entity, which can be clicked to search. (5) Entities show up as different colors and shapes of nodes based on their type and properties.}
    \Description{Lagoon UI screenshot}
    \label{fig:ui}
\end{figure}

An example of the \ac{UI} is shown in Fig. \ref{fig:ui}. To lay out the entities and observations, Cytoscape.js \cite{franz2016cytoscape} was used. To deal with the uniquely large graph databases for \ac{OSS} communities, time filtering is applied. When an entity has observations outside of the current time window, two special nodes are created, with the most recent preceding and following observation times, to allow for crawling the complete graph without memorizing times of interest.

Currently, the \ac{UI} supports a limited family of runtime plugins which can be implemented outside of the core \ac{LAGOON} code. These plugins can add additional details to an entity being viewed. This has been used for, e.g., providing user access to \ac{ML} models.

\subsection{Analysis Scripts and Reports}
The \ac{LAGOON} platform includes a library for querying the underlying database. Example usage includes querying the commit density and neighboring subgraph of each contributing community member at different points in time, and subsequently training an \ac{ML} model to predict future disengagement from the project based on current events. \Ac{LAGOON} is also packaged with scripts which generate reports on collaborations amongst contributors, as a way of better understanding interactions within the community.

\section{Accessing and running \acs{LAGOON}}\label{sec:run}
\ac{LAGOON} is available as a public repository at \blueurl{https://github.com/GaloisInc/SocialCyberLAGOON}. \emph{Users are strongly encouraged to refer to the \code{README.md} file for complete guidance on operating \ac{LAGOON} and making it useful for other projects}. Generally, all commands may be ran through ``\code{./lagoon\_cli.py}'', which supports tab completion and documentation via the ``\texttt{typer}'' package \cite{typer}.

\begin{figure}[tbp]
  \centering
  \includegraphics[width=0.82\linewidth]{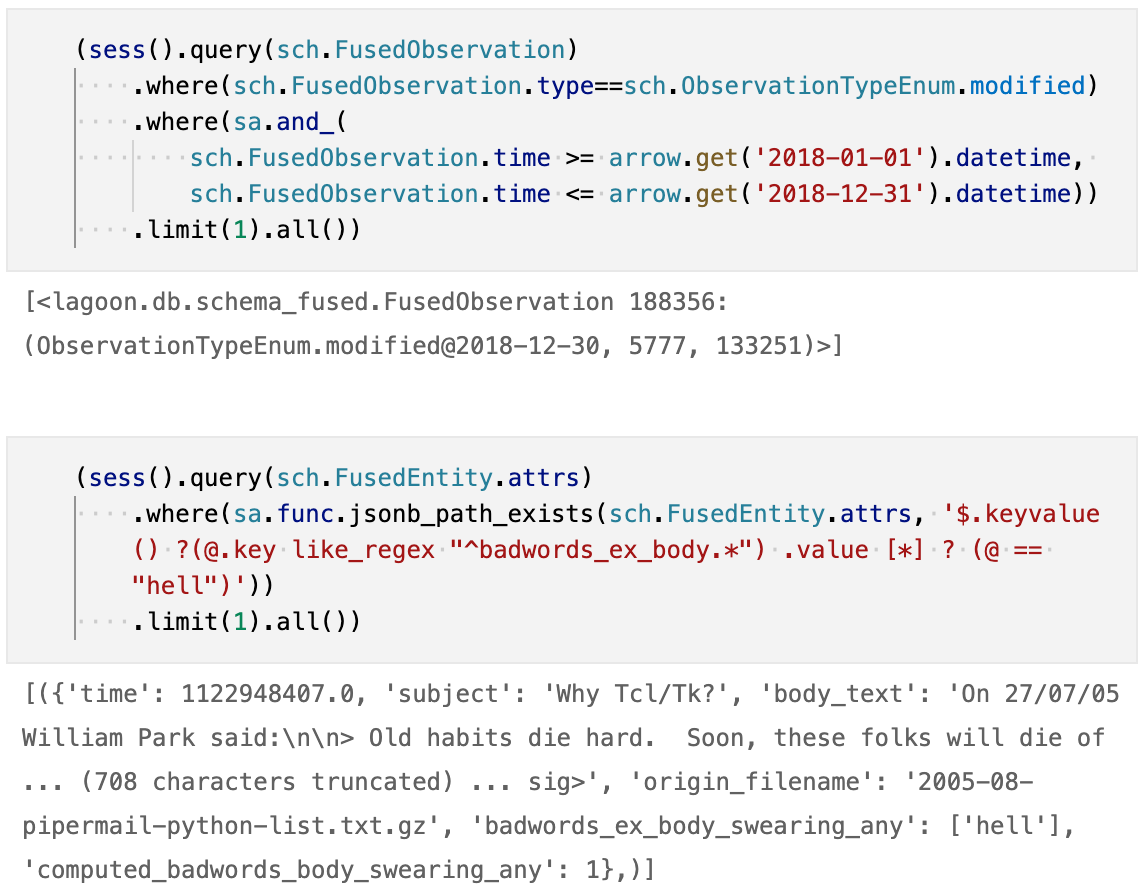}
  \caption{Two example database queries and their outputs (limited to one result for ease of display). Top -- Find observations made in 2018 that modified files. This is an example of time filtering. Bottom -- Find messages that contain the term `hell'. This is an example of detecting toxicity.}
  \Description{Example database queries}
  \label{fig:example_queries}
\end{figure}

As mentioned in the README, pre-populated databases that we have used can be downloaded from a publicly shared folder. Once the database is set up, the user can run ``\code{./lagoon\_cli.py shell}'' to interact with the database in an IPython environment using SQL queries. For example, ``\code{sess().query(sch.FusedEntity).first()}'' will print out the first entity. Fig. \ref{fig:example_queries} shows other example queries and their outputs.

\section{Our analyses on \acs{LAGOON} data}\label{sec:ana}
\begin{figure}[tbp]
  \centering
  \includegraphics[width=0.82\linewidth]{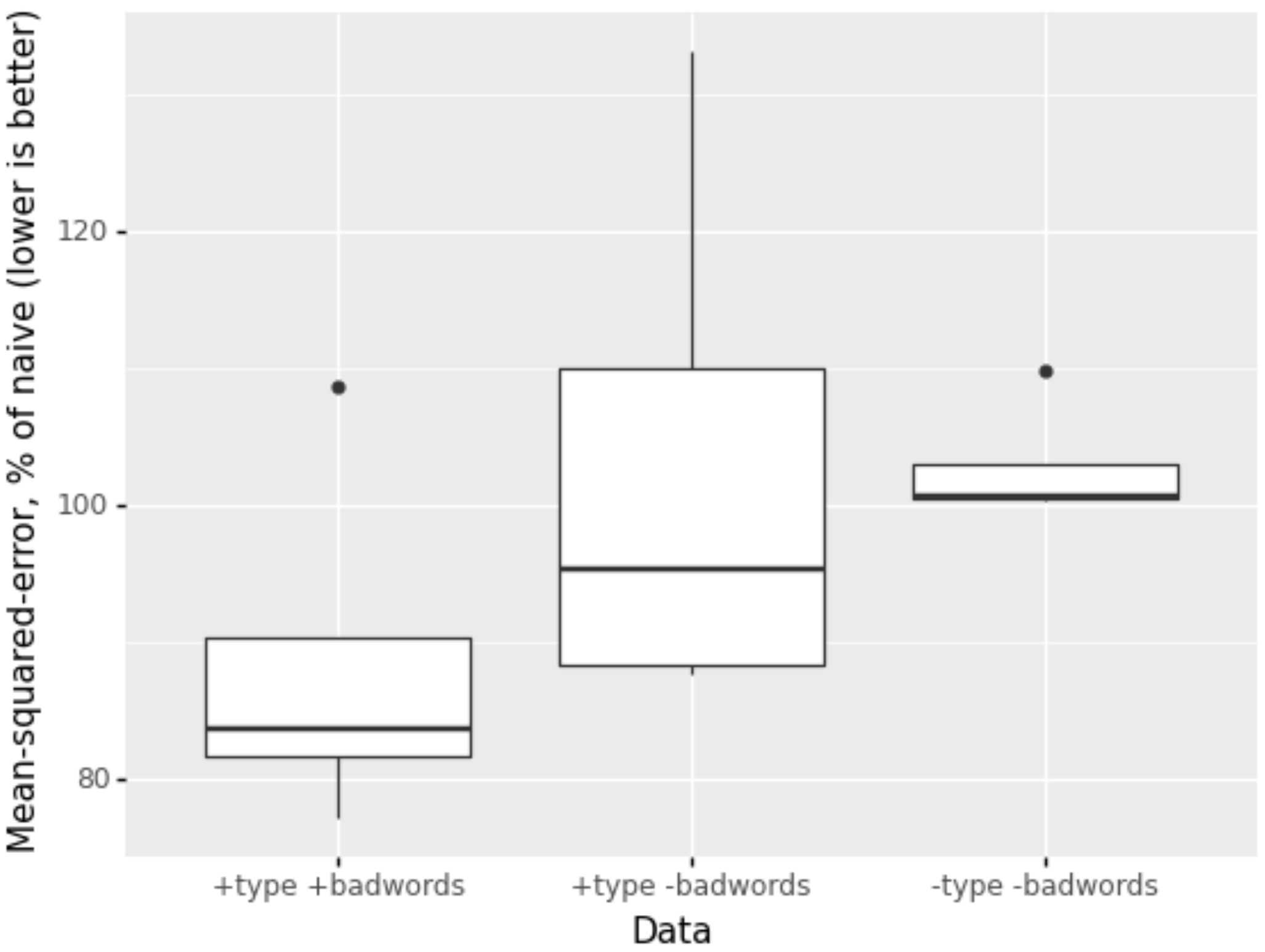}
  \caption{Prediction quality across varying levels of data inclusion. Numbers below $100\%$ indicate effective generalization from learning to contributors for which no instability information was included in the training data. Overall, including type and bad word counts as a proxy for toxicity led us to outperform a naive prediction of disengagement by $11.7\%$ on average; outlier detection in the box plot demonstrates a greater improvement when considering the median.}
  \Description{Results from the transformer enhanced GCN}
  \label{fig:gcn_results}
\end{figure}

We used a \ac{GCN} \cite{Kipf2017} enhanced with a transformer \cite{Vaswani2017} to predict contributor disengagement. This approach filtered the batch of toxicity attributes attached to messages to focus on those linked to a particular contributor (spatial analysis) over a particular period of time (temporal analysis). Disengagement was defined as the reduction in a contributor's activity from one time window to the next. Combining toxicity with vector embeddings for each type of entity in the graph led to an $11.7\%$ improvement over a naive predictor, as shown in Fig. \ref{fig:gcn_results}. Additional analyses, such as on \acp{PEP}, might be presented in future work, but are out of scope for this paper.

\section{Conclusion}\label{sec:conc}
We have presented \ac{LAGOON} -- an open source, reusable tool for analyzing \ac{OSS} communities. The key highlights include various ingestion modules, data layering, entity fusion, a \ac{UI}, and predictions on the health of \ac{OSS} communities.

The future of \ac{LAGOON} comprises adding more ingestion modules and increasing the power of the platform to predict threats to \ac{OSS} communities. The authors are happy to accept pull requests and extensions to \ac{LAGOON}, and hope that the broader community can benefit from using it.

\begin{acks}
This material is based upon work supported by the Defense Advanced Research Projects Agency (DARPA) under Contract No. HR00112190092. Any opinions, findings and conclusions or recommendations expressed in this material are those of the author(s) and do not necessarily reflect the views of the Defense Advanced Research Projects Agency (DARPA).
\end{acks}

\bibliographystyle{ACM-Reference-Format}
\bibliography{aaa_main}



\end{document}